\documentstyle[preprint,aps]{revtex}
\begin{document}
\title{An interpolation between Bose, Fermi and Maxwell-Boltzmann 
statistics based on Jack Polynomials}
\author{S. Chaturvedi, and V. Srinivasan}
\address{School of Physics\\
University of Hyderabad\\
Hyderabad - 500 046 (INDIA)}
\maketitle
\begin{abstract}
An interpolation between the canonical partition functions of Bose, Fermi and 
Maxwell-Boltzmann statistics is proposed. This interpolation makes use of the 
properties of Jack polynomials and leads to a physically appealing 
interpolation between the statistical weights of the three statistics. This,
in turn, can be used to define a new exclusion statistics in the spirit of the 
work of Haldane.
\end{abstract}
\newpage

The exclusion statistics, introduced by Haldane in a pioneering work$^{1}$, 
has attracted considerable interest in recent years$^{2-4}$. This statistics, 
formulated without any reference to spatial dimensions, provides an 
interpolation between Bose and Fermi statistics and captures the essential 
features of the anyon or fractional statistics$^{5}$ peculiar to  
two dimensional systems. The initial motivation for introducing 
such a statistics incorporating the notion of a generalized exclusion 
principle came from the properties of quasi particles in the fractional 
quantum Hall effect and in one dimensional inverse-square exchange spin 
chains$^{6}$. The relevance of this statistics for  quasi 
particles of the one dimensional ideal gas described by the Calogero-
Sutherland model$^{7}$ has also been recently demonstrated$^{8}$.  
In its simplest form, Haldane statistics, rests on the observation 
that in an assembly of $N$ particles each of which can occupy $M$ states, the 
number of states available to the $N^{th}$ particle after $N-1$ particles have 
been distributed into various states is equal to $M$ for a Bose system and 
$M-(N-1)$ for a Fermi system. One can thus associate an effective dimension 
$d_{N}^B = M$ with a Bose system and $d_{N}^F = M-(N-1)$ with a Fermi system 
and, by way of generalization, look 
upon the two statstics as special cases of a generalized statistics 
characterized an effective dimension $d_{N}^g = M-g(N-1)$. The signficance of 
the concept of an effective dimension lies in the fact that it not only 
reflects the nature of the statistics through $g$, taken to be a measure of 
the statistical interactions, but also enables one to unify the statistical 
weights associated with Bose and Fermi systems through the following formula  
\begin{equation}
W_g = \left( \begin{array}{c} d_{N}^g +N-1 \\ N \end{array} \right)
=  \left( \begin{array}{c} M+(1-g)(N-1) \\ N \end{array} \right) \,\,\,,
\end{equation} 
which for $g= 0$ and $g=1$ clearly reduces to the familiar expressions for the 
statistical weights $W_B$ and $W_F$ associated with Bose and Fermi systems 
respectively. Thermodynamic properties of an ideal generalized gas specified 
by the statistical weight given in $(1)$ have been investigated by a number of 
authors$^{2-4}$.

The interpolation between Bose and Fermi statistics furnished by $(1)$ is at 
the level of the total number of quantum states of $N$ particles corresponding 
to Bose and Fermi statistics. It is therefore natural to seek an interpolation 
between the two at a finer level i.e. at the level of canonical partition 
functions. In this letter we propose one such interpolation based on Jack 
polynomials$^{9}$. A signficant feature of this interpolation is that it 
not only encompasses Bose and Fermi statistics but also the "corrected" 
Maxwell-Boltzmann statistics. This interpolation between the canonical 
partition functions of the three standard ideal gas statistics, in turn,
leads to an interpolation between their statistical weights which, like 
$(1)$, can be used to define an exclusion statistics in the spirit of 
Haldane's work.

In a recent work$^{10}$ it was shown that the Schur functions$^{11}$  
$s_{\lambda}(x);~ x \equiv (x_1,\cdots,x_M)$, a class of symmetric 
polynomials in $x_1,\cdots,x_M$ of degree $N$ indexed by the partitions 
$\lambda$ of $N$, occupy a privileged position in all quantum statistics based 
on the permutation group . In particular, it was shown that, with the 
identification $x_i = e^{\beta \epsilon_i}$, $\epsilon_i$ being the single 
particle energies, the canonical partition functions for Bose and Fermi 
systems can be expressed in terms of a single Schur function correponding to 
the partitions $\lambda =(N,0,0, \cdots)$ and $\lambda =(1^N,0,0, \cdots)$ 
respectively
\begin{equation}
Z_{N}^B (x) = s_{(N)}(x) \,\,\,,
\end{equation}
\begin{equation}
Z_{N}^F (x) = s_{(1^N)}(x) \,\,\,.
\end{equation}
The canonical partition function for the "corrected" Maxwell-Boltzmann 
statistics is given by the familiar expression
\begin{equation}
Z_{N}^{MB} (x) = (x_1+ x_2 +\cdots+x_M)^N /N!
\end{equation}
with the $N!$ in the denominator accounting for the "correction". We would 
like to emphasize here that the observation that the the canonical partition
functions for Bose and Fermi systems can be expressed as in $(1)$ and $(2)$ 
plays a crucial role in the unification of the three canonical partition 
functions given above in terms of Jack polynomials.

Jack polynomials denoted by $J_{\lambda}(x;\alpha)$, like the 
Schur functions, constitute  a class of symmetric polynomials in 
$x_1,\cdots,x_M$ of degree $N$ indexed by the partitions $\lambda$ 
of $N$ and depend on a parameter $\alpha$. These polynomials, though studied 
extensively in the mathematical literature$^{9}$, have acquired prominence in 
the physics literature only recently through the fact that they appear in the 
eigenfunctions of the Calogero-Sutherland Hamiltonian. Some of their 
properties relevant to the present work are summarised below

\begin{itemize}
\item[[1]] $J_\lambda (x;\alpha)$ if normalized in such a way that when 
expressed in terms of the monomial symmetric functions $m_\kappa (x)$, 
\begin{equation}
J_\lambda (x;\alpha) = \sum_{\kappa} c_{\lambda \kappa} (\alpha) 
m_\kappa (x)\,\,\,,
\end{equation}
the coefficient of $m_{(1^N)}(x)$ in this expansion is $N!$, then 
$c_{\kappa \lambda}(\alpha)$ are polynomials in $\alpha$ with non negative 
integer coefficients. The monomial symmetric functions introduced above are
defined as follows$^{11}$.
\begin{equation}
m_\lambda(x) = \sum x_1^{\lambda_1}x_2^{\lambda_2} \cdots x_M^{\lambda_M}
\,\,\,\,.
\end{equation}
The sum on the rhs of $(6)$ is over all  distinct permutations of 
$(\lambda_1,\cdots,\lambda_M)$.
\item[[2]] For $\lambda = (N,0,0,\cdots)$ the coefficients of expansion in
$(5)$ are explicitly known$^{12}$ and one has
\begin{equation}
J_{(N)} (x;\alpha) = \sum_{\lambda}\frac{N!}{{\lambda_1}!\cdots {\lambda_M}!}
\prod_{i}[1.(1+\alpha)(1+2\alpha)\cdots(1+(\lambda_i -1)\alpha) ]~~ 
m_\lambda (x)
\,\,\,.
\end{equation}
\item[[3]] The value of $J_{(N)}(x;\alpha)$ for $x_1=x_2=\cdots x_M =1$ is 
given by
\begin{equation}
J_{(N)}(1;\alpha) = M(M+\alpha)(M+2\alpha)\cdots (M+(N-1)\alpha) \,\,\,.
\end{equation}
\end{itemize}

Setting $\alpha=1,~0,~-1$ in $(7)$, and using 
\begin{equation} 
\sum_{\lambda} m_{\lambda}(x) =s_{(N)}(x)\,\,\,, 
\end{equation}
\begin{equation}
\sum_{\lambda}\frac{N!}{{\lambda_1}!\cdots 
{\lambda_M}!}~m_\lambda (x)=(x_1+ x_2 +\cdots+x_M)^N ~~~~(multinomial~theorem)
\,\,\,,
\end{equation}
\begin{equation}
m_{(1^N)}(x) = s_{(1^N)}(x)
\end{equation}
respectively
one obtains 
\begin{equation}
J_{(N)} (x;1) = N!~s_{(N)}(x)~~;~~J_{(N)} (x;0) = (x_1+\cdots,x_M)^N 
~~;~~J_{(N)} (x;-1) = N!~s_{(1^N)}(x)~~.
\end{equation}
Hence if we define
\begin{equation}
Z_{N}^{\alpha} (x) = J_N(x.\alpha) /N! \,\,\,,
\end{equation}
then, in view of $(2)-(4)$, we have
\begin{equation}
Z_{N}^{\alpha} (x)~~~ =~~~~ \begin{array}{c} Z_{N}^B (x)~~~~if~\alpha=1 \\ 
Z_{N}^{MB}(x)~~~if~\alpha=0 \\ ~Z_{N}^{F}(x)~~~~~if~\alpha=-1 \end{array}
\end{equation}
Thus we find that the function defined in $(13)$ interpolates between the 
canonical partition functions of the three standard statistics. Setting 
$x_i = 1$ and using $(8)$, one finds that the corresponding statistical 
weight $W_\alpha $ 
\begin{equation}
W_\alpha \equiv Z_{N}^\alpha (1) = \frac{M(M+\alpha)(M+2\alpha)\cdots
(M+(N-1)\alpha)}{N!} \,\,\,,
\end{equation} 
as expected, interpolates between the statistical weights 
\begin{equation}
W_B = \left( \begin{array}{c} M+N-1 \\ N \end{array} \right)~~~;~~~
W_{MB} = \frac{M^N}{N!}~~~;~~~
W_{F} = \left( \begin{array}{c} M \\ N \end{array} \right)\,\,\,,
\end{equation}
associated with Bose, Maxwell-Boltzmann and Fermi statistics. If we introduce,
a l\'{a} Haldane, a concept of an effective dimension through $d_{N}^\alpha 
= M + \alpha (N-1)$, then $(15)$ may be written as 
\begin{equation}
W_\alpha = \frac{1}{N!} \prod_{i}^{N} d_{i}^\alpha\,\,\,.
\end{equation}
The interpolations furnished by $(1)$ and $(15)$ are clearly different. 
However, if we put $\gamma =1/\alpha$ in $(15)$, we find that it can be cast 
into the form
\begin{equation}
W_\alpha = \frac{1}{\gamma^{N}} \left( \begin{array}{c} \gamma M+N-1 \\ N 
\end{array} \right) \,\,\,, 
\end{equation}
similar to that in $(1)$. 

Thus we find that $Z_{N}^\alpha $ given in $(13)$ has a nice feature that it 
not only interpolates between the canonical partition functions of the three 
standard statistics but also leads to a physically appealing formula $(17)$ 
for the statistical weight $W_\alpha $. The statistical weight is simply given 
by the product of the effective dimensions divided by $N!$. The picture 
underlying the effective dimensions is also consistent with physics. 
Statistical interactions owe their origin to  the symmetry requirements 
imposed on the $N$ particle wave function. The deviations from the ideal gas 
law $PV=NkT$ provide a quantitative measure of their strength and nature. 
In Fermi systems the, the anti symmetrization of the wavefunction leads to a 
repulsive statistical interaction and one expects the effective dimension to 
decrease with $N$. In Bose systems, symmetrization of the wavefunction 
gives rise to an attractive statistical interactions and one expects the 
associated effective dimension to increase with $N$. As for Maxwell-Boltzmann 
statistics, since no notion of symmetrization or anti symmetrization is 
involved and since the ideal gas law always holds, there are no statistical 
interactions and one expects the effective dimension to be independent of $N$. 
This is precisely what is borne out by the assignment of effective dimensions 
to the three statistics given above. One can thus define a generalized 
statistics based on $(17)$ with $\alpha$ as a measure of the statistical 
interactions consistent with the intuitive picture one has about the origin 
of these interactions. Thermodynamic consequences of this generalized 
statistics based on $(17)$ can be worked out in much the same way as has been 
done for the one based on $(1)$. The analysis turns out to be much simpler 
and one finds that, for arbitrary $\alpha$, the mean number of particles 
$\overline {N_i}$ in the state $i$ is given by   
\begin{equation}
\overline{N_i} = \frac{1}{e^{\beta (\epsilon_i-\mu)}\pm{\mid \alpha\mid}} 
\,\,\,.
\end{equation}
where the positive or the negative sign on the rhs is to be taken depending 
$\alpha$ is negative or positive, and in the two cases 
one may speak of a Bose-like or a Fermi-like statistics. The chemical 
potential $\mu$ that enters $(19)$ is to be determined by requiring that 
$\overline{N} =\sum_i \overline{N_i}$. From the structure of $(19)$ 
it is clear that the expressions for various thermodynamic quantities 
appropriate to this statistics can simply be obtained by replacing 
$\overline{N}$ by $\overline{N}/{\mid \alpha \mid }$ in the corresponding 
expressions for Bose and Fermi statistics. Thus, for instance, the second 
virial coefficient for an ideal Bose-like or Fermi-like gas will simply be 
$\mid \alpha \mid $ times that for Bose and Fermi systems. 

To compare the exclusion statistics based on $(1)$ and $(17)$, consider 
the cases corresponding to $g=1/2$ in $(1)$ and  $\alpha= -1/2 $ in $(17)$. 
For $g=1/2$, though the explicit expression for $\overline{N_i}$ as a function 
of temperature turns out to be different from that in $(19)$, one finds that 
at $T=0$ both lead to the result that $\overline{N_i}= 2$ if 
$\epsilon_i < \mu$ and equal to zero if $\epsilon_i > \mu$ indicating 
fractional exclusion attributable to statistical interactions. However, when 
one examines the limit of high temperatures and low densities one finds that 
in the statistics based on $(1)$ the first correction to the ideal gas 
law turns out to be proportional to  $(2g-1)$ and, therefore, vanishes at 
$g=1/2$ suggesting absence of statistical interactions. In the statistics 
based on $(17)$, on the other hand, the effects of statistical interactions do 
show up in the first correction to the ideal gas law and magnitude of the 
correction is precisely half of that for a Fermi gas.        

To conclude, we have found a formula based on Jack polynomials which 
interpolates between the canonical partition functions of Bose, 
Fermi, and the corrected Maxwell- Boltzmann statistics. It also leads to 
a unifying expression for the statistical weight which can be given a simple 
physical interpretation on the basis of the notion of an effective dimension 
introduced by Haldane. This aspect of the interpolation presented here is 
important because, from a purely mathematical point of view, one can always 
write any number of formulae which would contain the three canonical partition 
functions as special cases. One such possibility could be 
\begin{equation}
Z_{N}^\alpha (x) = \frac{1}{2} \alpha(\alpha+1) Z_{N}^B (x) 
+ (1- \alpha)(\alpha+1) Z_{N}^{MB} (x)
+ \frac{1}{2} \alpha(1-\alpha) Z_{N}^F (x)
\end{equation}
The corresponding formula for the statistical weights would, however, 
be devoid of any intuitive appeal. It is also remarkable the generalized 
statistics presented here automatically contains the "corrected" 
Maxwell-Boltzmann statistics (as opposed to the infinite or uncorrected 
Maxwell-Boltzmann statistics) as a special case. We would also like to note 
that, for specific values of $\alpha$, Jack polynomials can be viewed as 
zonal spherical functions on certain symmetric spaces and this perspective 
may lead to a deeper understanding of the generalized statistics which goes 
beyond simple combinatorial considerations.
 
\vskip3.0cm 
\noindent{\bf Acknowledgements}

We are grateful to Dr. P.K. Panigrahi for numerous dicussions.

\newpage
\noindent{\bf References}
\begin{enumerate}
\item F.D.M. Haldane, Phys.Rev.Lett. {\bf 67}, 937 (1991).
\item Yong-Shi Wu, Phys.Rev.Lett. {\bf 73}, 922 (1994).
\item C. Nayak and F. Wilczek, Phys.Rev.Lett. {\bf 73}, 2740 (1994).
\item A.K. Rajgopal, Phys.Rev.Lett. {\bf 74}, 1048 (1995).
\item J.M.Leinaas and J. Myerheim, Nuovo Cimento {\bf 37}B, 1 (1977); 
F. Wilczek, Phys.Rev.Lett. {\bf 49}, 957 (1982); see also 
{\it Fractional statistics and anyon superconductivity}; edited by 
F. Wilczek (World Scientific,Singapore,1992), and references therein.
\item F.D.M. Haldane, Phys.Rev.Lett. {\bf 66}, 1529 (1991).
\item F. Calogero, J. Math. Phys. {\bf 10}, 2191 (1969); {\bf 10}, 2197 
(1969); B. Sutherland J. Math. Phys. {\bf 12}, 246 (1971); J. Math. Phys. 
{\bf 12}, 251 (1971); Phys.Rev. A {\bf 4}, 2019 (1971); A {\bf 5}, 
1372 (1972)
\item M.V.N. Murthy and R. Shankar Phys.Rev.Lett. {\bf 73}, 3331 (1994);
S.B.Isakov Phys.Rev.Lett. {\bf 73}, 2150 (1994); G. Bernard and Yong-Shi Wu 
University of Utah Report No. UUHEP/94-03.
\item H.Jack Proc.Roy.Soc (Edinburgh) {bf 69}A, 1 (1970); I.G. Macdonald in 
{\it Algebraic groups Utrecht 1986}, edited by A.M. Cohen, W.H. Hesselink, 
W.L.J. van der Kallen and J.R. Strooker Lecture notes in Math {\bf 1271}, 189 
(1987) (Springer-Verlag, Berlin, Heidelberg, New york, 1987)
\item S.Chaturvedi, Phys.Rev. E {\bf ??}, ??? (1996)
\item See, for instance, I.G. Macdonald, {\it Symmetric functions and Hall  
polynomials} (Clarendon, Oxford, 1979).
\item R.P. Stanley Adv. in Math. {\bf 77}, 77 (1989).
\end{enumerate}
\end{document}